\pdfoutput=1

\documentclass[aps,prd,preprint,superscriptaddress,tightenlines,nofootinbib]{revtex4}

\usepackage{epsfig}
\usepackage{graphicx}
\usepackage{dcolumn}
\usepackage{bm}
\usepackage{rotating}

\newcommand{\mevcc}{\!\mathrm{MeV}/c^2}
\newcommand{\mevc}{\!\mathrm{MeV}/c}
\newcommand{\mev}{\!\mathrm{MeV}}
\newcommand{\gevcc}{\!\mathrm{GeV}/c^2}
\newcommand{\gevc}{\!\mathrm{GeV}/c}
\newcommand{\gev}{\!\mathrm{GeV}}

\begin{document}

\preprint{CLNS 08/2034}       
\preprint{CLEO 08-17}         

\title{\boldmath Observation of $\chi_{cJ}$ radiative decays to light vector mesons}

\author{J.~V.~Bennett}
\author{R.~E.~Mitchell}
\author{M.~R.~Shepherd}
\affiliation{Indiana University, Bloomington, Indiana 47405, USA }
\author{D.~Besson}
\affiliation{University of Kansas, Lawrence, Kansas 66045, USA}
\author{T.~K.~Pedlar}
\affiliation{Luther College, Decorah, Iowa 52101, USA}
\author{D.~Cronin-Hennessy}
\author{K.~Y.~Gao}
\author{J.~Hietala}
\author{Y.~Kubota}
\author{T.~Klein}
\author{B.~W.~Lang}
\author{R.~Poling}
\author{A.~W.~Scott}
\author{P.~Zweber}
\affiliation{University of Minnesota, Minneapolis, Minnesota 55455, USA}
% ...
\collaboration{CLEO Collaboration}
\noaffiliation

\date{July 23, 2008}

\begin{abstract} 
Using a total of 2.74 $\times 10^7$ decays of the $\psi(2S)$ collected with the CLEO-c detector, we present a study of $\chi_{cJ}\to\gamma V$, where $V=\rho^0,\omega,\phi$.  The transitions $\chi_{c1}\to\gamma\rho^0$ and $\chi_{c1}\to\gamma\omega$ are observed with $\mathcal{B}(\chi_{c1}\to\gamma\rho^0) = ( 2.43 \pm 0.19\pm 0.22 )\times 10^{-4}$ and $\mathcal{B}(\chi_{c1}\to\gamma\omega) = ( 8.3 \pm 1.5 \pm 1.2 )\times 10^{-5}$.  In the $\chi_{c1}\to\gamma\rho^0$ transition, the final state meson is dominantly longitudinally polarized.  Upper limits on the branching fractions of other $\chi_{cJ}$ states to light vector mesons are presented.
\end{abstract}

\pacs{13.20.Gd, 14.40.Gx}
\maketitle

Radiative decays of charmonium provide a rich context in which the interplay between theory and experiment can advance our understanding of quantum chromodynamics (QCD).  The radiative decays of the $J/\psi$ that proceed through annihilation of the $c\bar{c}$ quarks are of particular interest for spectroscopy as they provide a gluon-rich hadronic system recoiling against the radiated photon.  Such experimental channels are thought to be ideal for searching for bound states of gluons (glueballs); however, in order to interpret experimental data for these decays, one must have an understanding of radiative transitions of $J/\psi$ to light $P$-wave isoscalar ($f_J$) states.  In the case of the scalars ($f_0$), the picture is complicated by the uncertainty in the structure and properties of the many observed experimental states.  The radiative decays of $P$-wave charmonium ($\chi_{cJ}$) to light quark vector states ($\rho^0$, $\omega$, and $\phi$) provide an independent, complementary, $c\bar{c}$-annihilation decay where the properties and structure of the final state hadronic system are well-known, which may be useful in validating theoretical techniques.

In this Letter, we present the first observation of radiative decays of the $\chi_{c1}$ to the light vector mesons $\rho^0$ and $\omega$.  The measured rates for these decays are an order of magnitude higher than those predicted by Gao, Zhang, and Chao~\cite{Gao:2006bc} with perturbative QCD (pQCD) methods.

The data used in this analysis were taken with the CLEO-c detector operating at the Cornell Electron Storage Ring (CESR)~\cite{cesr}, which provided symmetric $e^+e^-$ collisions at the $\psi(2S)$ center-of-mass.  The detector, described in detail elsewhere~\cite{cleodet,cesr}, features a solid angle coverage of $93\%$ for charged and neutral particles. The charged particle tracking system operates in a $1.0~\!\mathrm{T}$ axial magnetic field and achieves a momentum resolution of $\approx\!0.6\%$ at $p=1~\gevc$. The CsI(Tl) calorimeter attains photon energy resolutions of $2.2\%$ at $E_{\gamma}=1~\gev$ and $5\%$ at $100~\mev$. Two particle identification systems, one based on ionization energy loss ($dE/dx$) in the drift chamber and the other a ring-imaging \v Cerenkov (RICH) detector, are used to identify pions, kaons, and protons. Detection efficiencies are determined using a \textsc{geant}-based~\cite{geant} Monte Carlo~(MC) detector simulation.

To enhance photon energy resolution and reduce background, photon candidates are required to be detected in the barrel portion of the calorimeter ($|\cos \theta| < 0.81$) and must be spatially separated from the trajectories of charged tracks that have been extrapolated to the calorimeter.  We form $\pi^0$ candidates from two photons whose invariant mass $M(\gamma\gamma)$ is less than three standard deviations from the nominal $\pi^0$ mass.  For charged particles, we require a hit in at least 50\% of the radial layers intercepted by the trajectory of the particle in the drift chamber, the $\chi^2/$d.o.f.\ for the fit to the hits be less than 50, and the charged particle be consistent with originating from the $e^+e^-$ interaction.  To reduce backgrounds from Bhabha events, we additionally require that $|\cos \theta|< 0.83$ for reconstructed charged tracks.  Defining $\sigma_{X}$ as the number of standard deviations the measured $dE/dx$ is away from the expected $dE/dx$ for a particle of type $X$, we identify pions and kaons by requiring $\sigma_\pi < 4$ and $\sigma_K < 4$, respectively.  In addition, we utilize information from the RICH detector:  $L_X \equiv -2 \ln \mathcal{L}_X$, where $\mathcal{L}_X$ is the likelihood that the signature in the RICH is from a particle species $X$.  For kaon candidates with $p>800~\mevc$ that produce a signal in the RICH detector, we require $L_K - L_\pi +\sigma^2_K - \sigma^2_\pi < 0$ and $L_K - L_p +\sigma^2_K - \sigma^2_p < 0$.

We reconstruct the exclusive decay $\psi(2S)\to\gamma_{\mathrm{l}}\chi_{cJ}$; $\chi_{cJ}\to\gamma_{\mathrm{h}}V$, where $\gamma_{\mathrm{l}}$ ($\gamma_{\mathrm{h}}$) designates the characteristic low (high) energy photon in the signal topology and $V$ is either a $\rho^0$, $\omega$, or $\phi$ candidate.  The $\rho^0$, $\omega$, and $\phi$ candidates are reconstructed in the $\pi^+\pi^-$, $\pi^+\pi^-\pi^0$, and $K^+K^-$ decay modes, respectively.  A four-constraint kinematic fit is performed to the entire event which forces the decay products to be consistent with the known four-momentum of the initial $\psi(2S)$.  Candidates that have  $\chi^2/$d.o.f.~$< 5$ for this fit are retained.  In the rare case that an event has more than one candidate, only the candidate with the smallest $\chi^2/$d.o.f.\ is kept.  The kinematically-fitted four-momenta of the decay products are used for subsequent analysis.  To suppress multi-body hadronic decays of the $\psi(2S)$ we require that $|M(\gamma_\mathrm{l}\gamma_\mathrm{h}) - M(\pi^0)| > 15~\mevcc$ and $|M(\gamma_\mathrm{l}\gamma_\mathrm{h}) - M(\eta)| > 25~\mevcc$.

In the search for $\chi_{cJ}\to\gamma\rho^0$ it is necessary to eliminate the copious background from $e^+e^-\to (e^+e^-~\mathrm{or}~\mu^+\mu^-)$ where the leptons are misidentified as $\pi^+\pi^-$ and radiated photons fake $\gamma_\mathrm{l}$ and $\gamma_\mathrm{h}$.  This background can be effectively eliminated by placing requirements on the opening angles of the two photon and pion candidates in the laboratory frame:  $-0.70 < \cos\theta_{\pi^+\pi^-} < 0.90$ and $| \cos\theta_{\gamma_\mathrm{l}\gamma_\mathrm{h}} | < 0.98$.  To further suppress backgrounds from Bhabha events, the total detected energy in the calorimeter is required to be less than 90\% of the center of mass energy.  An additional background arises from decays of the type $\psi(2S)\to\gamma_\mathrm{h}\eta^\prime;\eta^\prime\to\gamma_\mathrm{l}\pi^+\pi^-$ and is suppressed by requiring $|M(\gamma_\mathrm{l}\pi^+\pi^-)-M(\eta^\prime)| > 15~\mevcc$.  

Our general strategy for extracting the signal is to select events using the invariant mass of the candidate vector meson and then plot the transition photon energy $E(\gamma_\mathrm{l})$ for events passing these selection criteria.  The signal for $\chi_{c0}$, $\chi_{c1}$, and $\chi_{c2}$ decay will appear as peaks in $E(\gamma_\mathrm{l})$.  The signal selection criteria for the three unique final states are $0.50 < M(\pi^+\pi^-) < 1.10~\gevcc$ ($\chi_{cJ}\to\gamma\rho^0$), $0.75 < M(\pi^+\pi^-\pi^0) < 0.82~\gevcc$ ($\chi_{cJ}\to\gamma\omega$), and $1.01 < M(K^+K^-) < 1.04~\gevcc$ ($\chi_{cJ}\to\gamma\phi$).

\begin{figure}
\begin{center}
\epsfig{width=\columnwidth, file=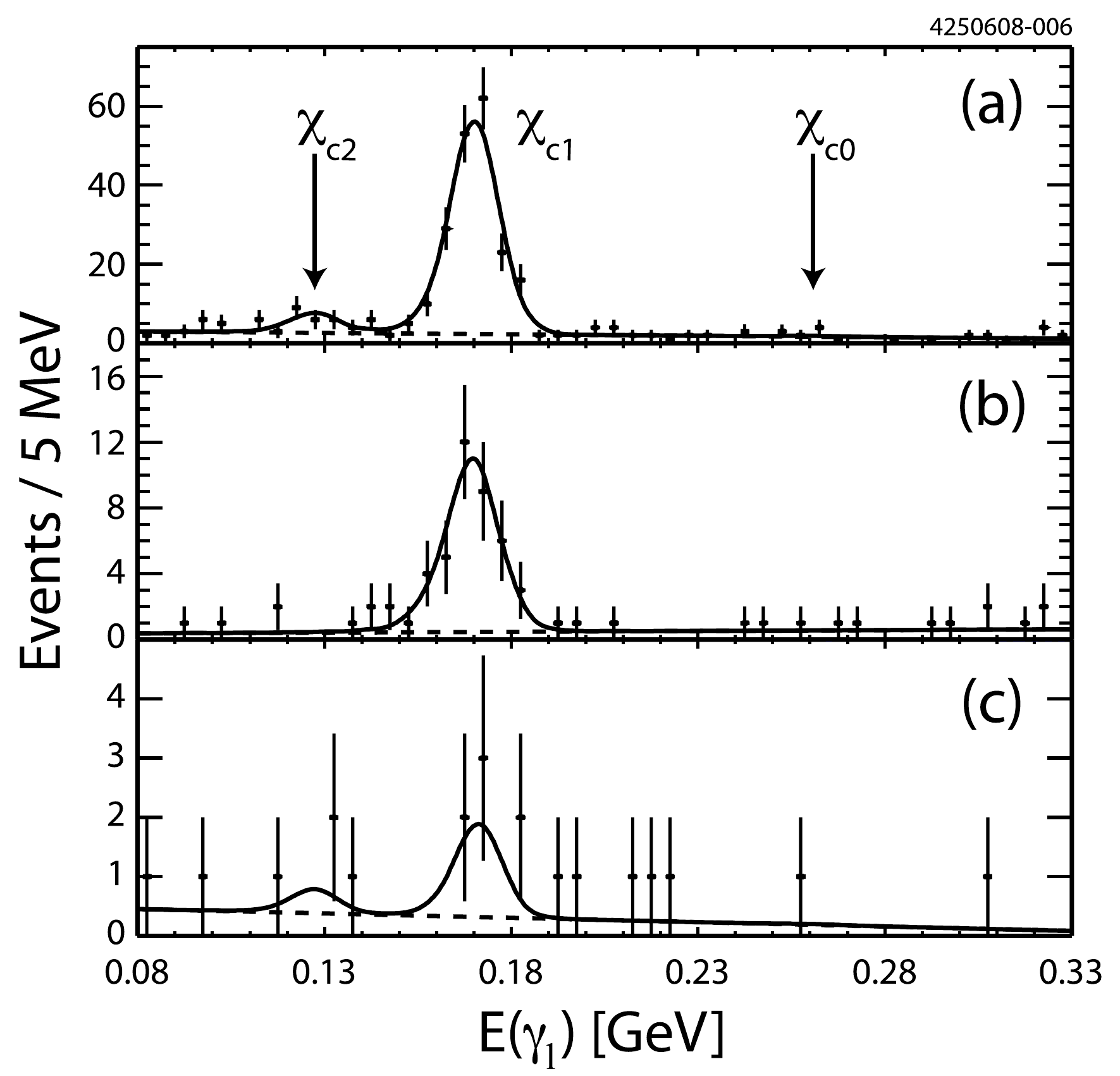}
\end{center}
\caption{The $\psi(2S)\to\gamma\chi_{cJ}$ transition photon ($\gamma_\mathrm{l}$) energy distribution for (a) $\chi_{cJ}\to\gamma\rho^0$, (b) $\chi_{cJ}\to\gamma\omega$, and (c) $\chi_{cJ}\to\gamma\phi$ candidates.  The data are shown by the points; the fit (described in the text) is shown as a solid line.  The background component of the fit is indicated by the dashed line.}
\label{fig:spectrum}
\end{figure}

The distribution of $\psi(2S)\to\gamma\chi_{cJ}$ transition photon energy $E(\gamma_\mathrm{l})$ is shown in Figs.~\ref{fig:spectrum}(a)-(c) for $\chi_{cJ}\to\gamma V$, where $V = \rho^0,\omega,~\mathrm{and}~\phi$, respectively.  Clear signals are observed for the $\chi_{c1}\to\gamma\rho^0$ and $\chi_{c1}\to\gamma\omega$ transitions.  To extract the event yield from the spectra we first obtain a signal shape for each of the nine $\chi_{cJ}\to\gamma V$ transitions using an MC simulation of the signal where the mass and full width of the $\chi_{cJ}$ are taken from Ref.~\cite{pdg}.  The MC simulation is subjected to the same kinematic fitting and analysis requirements as the data.  Each of the distributions in Fig.~\ref{fig:spectrum} is fit to a linear background shape and a sum of three signal shapes, one for each of the $\chi_{cJ}$ states.  The two parameters that describe the background and the normalization for each of the $\chi_{cJ}$ photon lines are allowed to float in the fit.  The fitted yields are summarized in Table~\ref{tab:results}. By examining the change in the fit likelihood when the signals yields are forced to zero, we estimate the significance of the $\chi_{c1}\to\gamma\rho^0$ and $\chi_{c1}\to\gamma\omega$ signals to be much greater than $5\sigma$, while the significance for $\chi_{c1}\to\gamma\phi$ is less than $3\sigma$.  These estimates do not
include systematic uncertainties, discussed below, that may affect the significance of the yield.

\begin{figure}
\begin{center}
\epsfig{width=\columnwidth, file=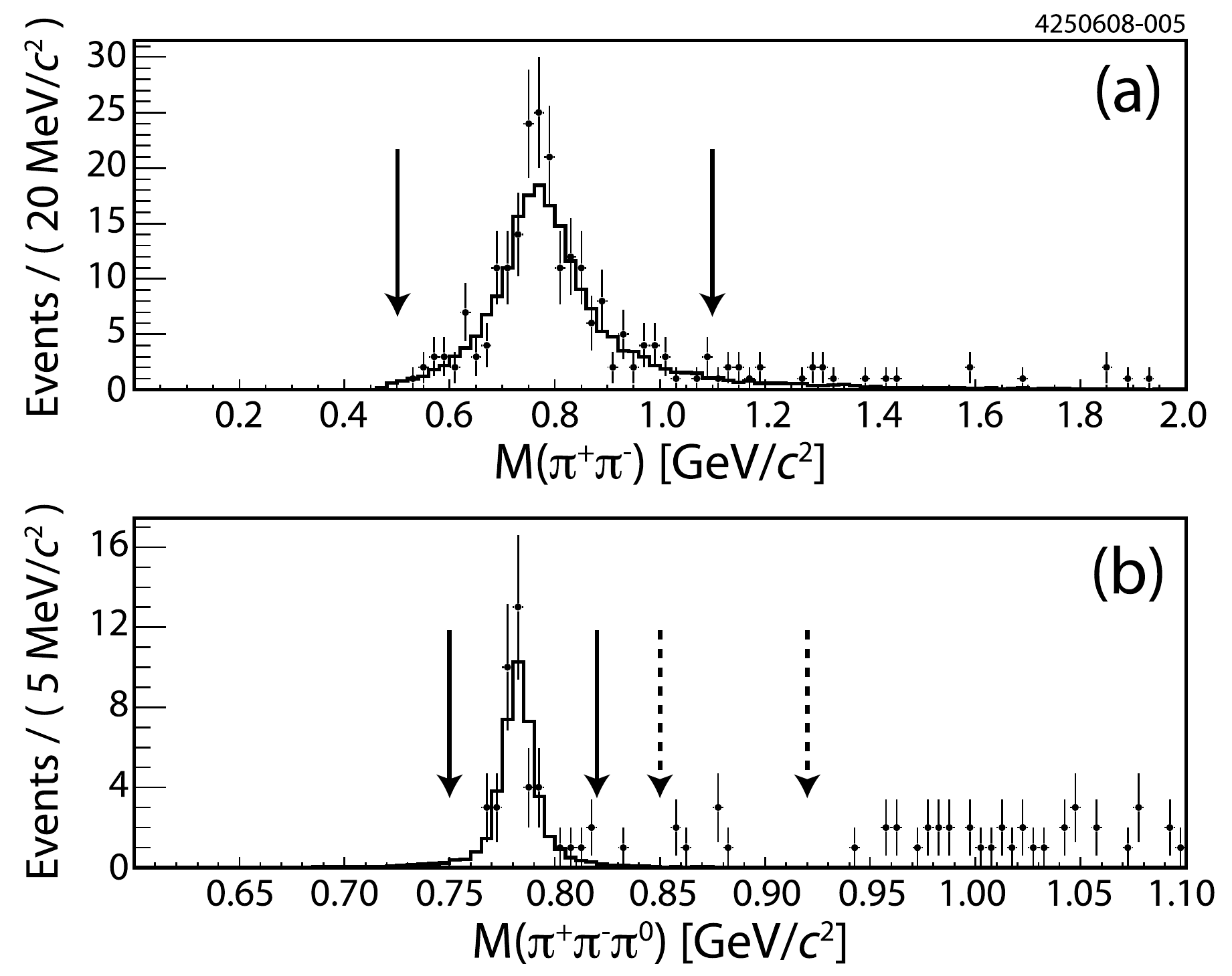}
\end{center}
\caption{Invariant mass of (a) $\rho^0$ and (b) $\omega$ candidates for events that pass all analysis criteria and have $E(\gamma_\mathrm{l})$ consistent with $\psi(2S)\to\gamma_\mathrm{l}\chi_{c1}$ ($150 < E(\gamma_\mathrm{l}) < 200$~MeV).  The points are data and the solid line is signal MC scaled to the yield extracted in the nominal fit.  The signal region is indicated by the solid arrows.  The sideband region for $\omega$ candidates (described in the text) is indicated by the dashed arrows.}
\label{fig:rho_omega}
\end{figure}

Our signal yield can be potentially biased by background from real $\chi_{cJ}$ decays, which peak in $E(\gamma_\mathrm{l})$, that are partially reconstructed thereby faking our signal.  Fortunately, hadronic decays of the type $\chi_{cJ}\to\pi^0+(\rho^0,\omega,~\mathrm{or}~\phi)$ are forbidden by $C$-parity conservation, otherwise they would certainly contribute a substantial peaking background to our signal.  Other hadronic decays such as $\chi_{cJ}\to K^+K^-\pi^0$ or $\chi_{cJ}\to\pi^+\pi^-\pi^0\pi^0$ are allowed.  In general, these either do not peak in vector meson invariant mass or require the loss of multiple neutral particles, and are consequently suppressed by the requirements placed on hadronic candidate invariant mass or $\chi^2$/d.o.f.\ of the kinematic fit.  In fact, using an MC simulation that models all $\psi(2S)$ and $\chi_{cJ}$ hadronic decays, we observe no such peaking backgrounds.  Figure~\ref{fig:rho_omega} shows the invariant mass distributions for $\rho^0$ and $\omega$ candidates in the $\chi_{c1}$ region of $E(\gamma_\mathrm{l})$ -- as is evident from the sideband regions, the bias due to non-$\rho^0$ or non-$\omega$ backgrounds is small.  Nevertheless, for those channels where we have sufficient statistics to do so, we adopt a data-driven approach to estimate this bias.  In the $\chi_{cJ}\to\gamma\rho^0$ case, we generate background-subtracted $E(\gamma_\mathrm{l})$ spectra by fitting the $\rho^0$ yield in bins of $E(\gamma_\mathrm{l})$.  Repeating this procedure with variations of the background parameterization in the $\rho^0$-candidate invariant mass spectrum resulted in a maximum deviation from the nominal efficiency-corrected yield of $-2\%$ ($-50\%$) for $\chi_{c1}(\chi_{c2})\to\gamma\rho^0$.  The nominal analysis was also repeated while altering the selected region in $M(\pi^+\pi^-)$.  Changes in the efficiency-corrected yield for the $\chi_{c1}(\chi_{c2})\to\gamma\rho^0$ signal ranged from -1\% to +2\% (-20\% to +20\%).  For $\chi_{c1}\to\gamma\omega$, we extract the yield from a fit to the $E(\gamma_\mathrm{l})$ spectrum obtained by selecting events in the $\omega$-candidate invariant mass sideband, $850 < M(\pi^+\pi^-\pi^0) < 920~\mevcc$ (shown in Fig.~\ref{fig:rho_omega}), and conservatively assume that this yield, $3.1\pm 3.2$ events (8\% of our signal yield), is equivalent to the background in the $\omega$ signal region in our nominal analysis.  In addition, we repeat the analysis for various selected regions in $M(\pi^+\pi^-\pi^0)$.  In both cases, changes in the efficiency-corrected yields for $\chi_{c1}\to\gamma\omega$ were never larger than $\pm8\%$.  In all cases described above, we find no statistically significant evidence for a bias in the efficiency-corrected yield.  The central values for the (insignificant) biases are used as a quantitative estimate of our uncertainty, summarized in Table~\ref{tab:results}.  In all other channels we conservatively estimate the upper limit on the rates by assuming that all observed events are signal.  

\begin{sidewaystable}
\caption{Summary of the fitted yield, efficiency, and branching fraction ($\mathcal{B}$) or upper limit (U.L.) at 90\% confidence level for each of the $\chi_{cJ}\to\gamma V$ transitions.  Also listed is the total systematic error and the portion of the systematic error due to uncertainty in the backgrounds that might bias the signal yield.  The efficiencies include the vector meson branching fractions~\cite{pdg} and the probability of detecting the $\psi(2S)\to\gamma\chi_{cJ}$ transition photon.  Finally, we list the pQCD predictions of Ref.~\cite{Gao:2006bc}.}
\label{tab:results}
\begin{tabular}{lccccccc}\\\hline\hline
Mode & Yield [Events] & Efficiency [\%] & Bias Uncert. [\%] & Syst. Error [\%] & $\mathcal{B}\times 10^{6}$  & U.L. $[10^{-6}]$  & ~~pQCD $[10^{-6}]$ \rule{0 pt}{11 pt} \\ \hline
$\chi_{c0}\to\gamma\rho^0$ & 1.2$\pm$4.5 & 30 & --  & $\pm10$ & ~ & $< 9.6$ & 1.2 \\
$\chi_{c1}\to\gamma\rho^0$ & 186$\pm$15 & 32 & $\pm 2$ &  $\pm 9$ & $243\pm19\pm 22$ & ~ & 14 \\
$\chi_{c2}\to\gamma\rho^0$ & 17.2$\pm$6.8 & 31 & $_{-50}^{+20}$ & $_{-57}^{+34}$ &  $25\pm10^{+8}_{-14}$ & $<50$ & 4.4 \\
$\chi_{c0}\to\gamma\omega$ &  0.0$\pm$2.8 & 17 & -- & $\pm16$ & ~ & $<8.8$ & 0.13 \\
$\chi_{c1}\to\gamma\omega$ & 39.2$\pm$7.1 & 20 & $\pm 8$ & $\pm 15$ & $83\pm15\pm12$ & ~ & 1.6 \\
$\chi_{c2}\to\gamma\omega$ & 0.0$\pm$1.8 & 18 & -- & $\pm 16$ & ~ & $<7.0$ & 0.50 \\
$\chi_{c0}\to\gamma\phi$ & 0.1$\pm$1.6 & 15 & -- & $\pm 12$ & ~ & $<6.4$ & 0.46 \\ 
$\chi_{c1}\to\gamma\phi$ & 5.2$\pm$3.1 & 17 & -- & $\pm 12$ & $12.8\pm7.6\pm1.5$ & $<26$ & 3.6 \\ 
$\chi_{c2}\to\gamma\phi$ & 1.3$\pm$2.5 & 16 & -- & $\pm 12$ & ~ & $<13$ & 1.1 \\ 
\hline\hline
\end{tabular}
\end{sidewaystable}

The efficiency for each mode (see Table~\ref{tab:results}) is obtained using an MC simulation that models the initial polarization of the $\psi(2S)$ and the appropriate electric-dipole (E1) angular distribution for the $\psi(2S)\to\gamma\chi_{cJ}$ transition photon.  The decay $\chi_{cJ}\to\gamma V$ is simulated uniformly in phase space except for the $\chi_{c1}\to\gamma\rho^0$ and $\chi_{c1}\to\gamma\omega$ decays. Here we modify the MC to reflect the measured polarization, described in detail below.
The efficiencies include branching fractions of the final state vector meson~\cite{pdg} and the detection efficiency for the initial transition photon $\gamma_\mathrm{l}$.

To obtain the product branching fractions $\mathcal{B}(\psi(2S)\to\gamma\chi_{cJ})\times\mathcal{B}(\chi_{cJ}\to\gamma V)$ we divide the yield by the product of the efficiency and the number of $\psi(2S)$ in our data sample, $2.74\times10^7$~\cite{npsi2s}.  The final $\chi_{cJ}\to\gamma V$ branching fractions are obtained by dividing by the appropriate E1 transition rate,  $\mathcal{B}(\psi(2S)\to\gamma\chi_{c0})=9.2\%$, $\mathcal{B}(\psi(2S)\to\gamma\chi_{c1})=8.7\%$, or $\mathcal{B}(\psi(2S)\to\gamma\chi_{c2})=8.1\%$~\cite{pdg}, and are summarized in Table~\ref{tab:results}.

In addition to the impact of the biases described above, we explore several other sources of systematic uncertainty.  Generous variations in the background parameterization used to fit the spectra in Fig.~\ref{fig:spectrum} produced variations no larger than 2\%, 25\%, and 2\% for the $\chi_{c1}\to\gamma\rho^0$, $\chi_{c2}\to\gamma\rho^0$, and $\chi_{c1}\to\gamma\omega$ yields, respectively.  Making significant changes in our event selection criteria produced variations in the efficiency corrected yield of 5\% and 8\% for the $\chi_{c1}\to\gamma\rho^0$ and $\chi_{c1}\to\gamma\omega$ channels, and we assign these respective systematic uncertainties to each of the $\chi_{cJ}\to\gamma\rho^0$ and $\chi_{cJ}\to\gamma\omega$ rates.  For $\chi_{cJ}\to\gamma\phi$, where we do not see a significant signal, we assume a systematic error due to event selection of 5\%, the same as $\chi_{c1}\to\gamma\rho^0$, which has a similar topology.  The errors in the track (photon) detection efficiency are assumed to be 1\% (2\%) per track (photon) and fully correlated across all tracks (photons).  The number of $\psi(2S)$ in our data sample is known with 2\%  precision~\cite{npsi2s}.  The uncertainty in polarization of the vector meson, assumed to be the maximum difference between phase space and either longitudinally or transversely polarized decays, introduces a 5\%, 10\%, and 8\% error for the $\chi_{cJ}\to\gamma(\rho^0,~\omega,~\mathrm{and}~\phi)$ efficiencies, respectively, with the exception of $\chi_{c1} \to\gamma\rho^0$ and $\chi_{c1}\to\gamma\omega$ modes where the polarization is measured and the resulting efficiency error due to uncertainty on this measurement is 1\% and 3\%, respectively.  All of the $\psi(2S)\to\gamma\chi_{cJ}$ rates have a relative 5\% uncertainty~\cite{pdg}.  The total systematic errors are summarized in Table~\ref{tab:results}.  Upper limits are scaled by $(1+\delta)$, where $\delta$ is the total relative systematic error.

\begin{figure}
\begin{center}
\epsfig{width=\columnwidth, file=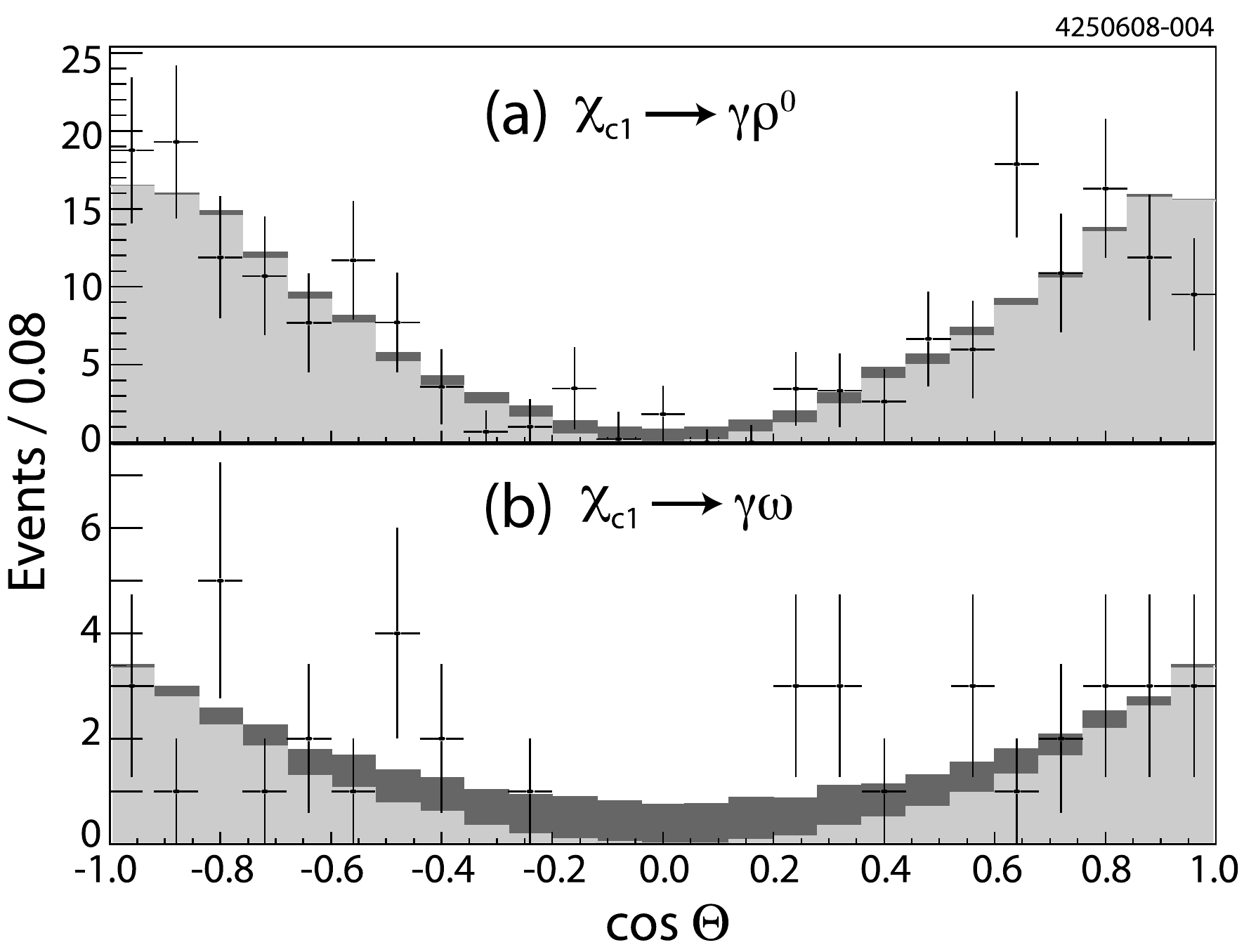}
\end{center}
\caption{Distributions in $\cos\Theta$ for the (a) $\chi_{c1}\to\gamma\rho^0$ and (b) $\chi_{c1}\to\gamma\omega$ candidates.  The histogram, a sum of longitudinal (light gray) and transverse (dark gray) components, shows the best fit to the data (points).}
\label{fig:pol}
\end{figure}

The ratio of transverse ($\lambda=\pm1$) to longitudinal ($\lambda=0$) polarization of the vector meson $A_\pm / A_0$ can be measured by examining the distribution of events as a function of $\cos\Theta$, where $\Theta$ is defined as the angle between the vector meson flight direction in the $\chi_{cJ}$ rest frame and either the $\pi^+$ direction in the $\rho^0$ rest frame or the normal to the decay plane in the $\omega$ rest frame.  Modulo detector acceptance, longitudinal (transverse) polarization exhibits a $\cos^2\Theta$ ($\sin^2\Theta$) dependence.  The distributions of $\cos\Theta$ are shown in Fig.~\ref{fig:pol}, where, for the $\chi_{c1}\to\gamma\rho^0$ case, the data (points) are obtained by fitting the invariant mass spectrum of the vector meson candidate in bins of $\cos\Theta$ in order to eliminate potential contamination from non-$\rho^0$ decays.  The $\chi_{c1}\to\gamma\omega$ candidates are plotted by requiring $150 < E(\gamma_\mathrm{l}) < 200$~MeV.  The individual transverse (dark gray) and longitudinal (light gray) components to which the data are fit are obtained from MC simulation, and the best fit, floating $A_\pm / A_0$  and the overall normalization, is indicated by the total solid histogram.  In principle, the decay amplitudes to the two polarization states can interfere; this interference is neglected in the fit.  The fits give $A_\pm / A_0 = 0.078^{+0.048+0.002}_{-0.036-0.022}$ for $\chi_{c1}\to\gamma\rho^0$ and $A_\pm / A_0 = 0.47^{+0.37+0.11}_{-0.24-0.23}$ for $\chi_{c1}\to\gamma\omega$, where the second, systematic error is obtained by assuming the estimated background contributes entirely to the longitudinal or transverse component.

In summary, we present the first observation of radiative decays of $\chi_{c1}$ to light vector mesons.  We find $\mathcal{B}(\chi_{c1}\to\gamma\rho^0) = ( 2.43 \pm 0.19\pm 0.22 )\times 10^{-4}$ and $\mathcal{B}(\chi_{c1}\to\gamma\omega) = ( 8.3 \pm 1.5 \pm 1.2 )\times 10^{-5}$.  The measured rates are significantly higher than those predicted by a calculation using pQCD~\cite{Gao:2006bc}, for which the leading-order decay mechanism is annihilation of the $c\bar{c}$ quarks into a light-quark pair that radiatively decays to $\gamma V$.   The longitudinally polarized structure of the $\chi_{c1}\to\gamma\rho^0$ decay parallels that measured in the decay of the corresponding light-quark axial-vector $f_1(1285)\to\gamma\rho^0$ by VES~\cite{Amelin:1994ii}.  This observation may suggest that the enhanced rate is due to the presence of a virtual light-quark axial-vector meson in the decay.  The branching fraction measurements and upper limits presented in this Letter provide input to cross-check current and future calculations of radiative decays of charmonia that are important for spectroscopic interpretations of experimental data.

We gratefully acknowledge the effort of the CESR staff in providing us with excellent luminosity and running conditions.  We thank T.~Barnes and J.~Dudek for interesting discussions.  This work was supported by the A.P.~Sloan Foundation, the National Science Foundation, the U.S. Department of Energy, the Natural Sciences and Engineering Research Council of Canada, and the U.K. Science and Technology Facilities Council.


\begin{thebibliography}{99}

\bibitem{Gao:2006bc}
  Y.~J.~Gao, Y.~J.~Zhang and K.~T.~Chao,
  Chin.\ Phys.\ Lett.\  {\bf 23}, 2376 (2006)
  [arXiv:hep-ph/0607278].
  
\bibitem{cesr} R.~A.~Briere {\it et al.} (CLEO-c/CESR-c Taskforces and CLEO-c Collaboration), Cornell University, LEPP Report No. CLNS 01/1742, 2001 (unpublished).

\bibitem{cleodet} Y.~Kubota {\it et al.} (CLEO Collaboration), Nucl. Instrum. Methods Phys. Res., Sect. A {\bf 320}, 66 (1992);  D.~Peterson {\it et al.}, Nucl. Instrum. Methods Phys. Res., Sect. A {\bf 478}, 142 (2002); M.~Artuso {\it et al.}, Nucl. Instrum. Methods Phys. Res., Sect. A {\bf 502}, 91 (2003).  

\bibitem{geant} R.~Brun {\it et al.}, GEANT 3.21, CERN Program Library Long Writeup W5013 (1993), unpublished.

\bibitem{pdg} W.-M.~Yao {\it et al.}, J. Phys. G {\bf 33}, 1 (2006) and 2007 partial update for 2008.

\bibitem{npsi2s} H. Mendez {\it et al.} (CLEO Collaboration), Phys.\ Rev.\ D {\bf 78}, 011102(R) (2008).

\bibitem{Amelin:1994ii}
  D.~V.~Amelin {\it et al.},
  Z.\ Phys.\  C {\bf 66}, 71 (1995).

\end{thebibliography}
\end{document}